\documentclass[journal]{vgtc}                
\pdfoutput=1

\usepackage{amssymb}
\usepackage{amsmath}
\usepackage{amsmath}

\usepackage{mathptmx}
\usepackage{graphicx}
\usepackage{times}
\usepackage{comment}
\usepackage{enumitem}
\usepackage{xspace}
\usepackage[binary-units]{siunitx}
\usepackage{algorithm}
\usepackage{algpseudocode}
\usepackage{listings}
\lstdefinestyle{myPy}{
language=Python,
numbers=left,
aboveskip=\smallskipamount,
belowskip=\smallskipamount,
stepnumber=1,
numbersep=3pt,
numberstyle=\tiny,
tabsize=2,
basicstyle=\ttfamily\tiny,
captionpos=b,
frame=tb,
escapeinside={(*@}{@*)},
  morekeywords={output,@property,None}
}
\usepackage[bookmarks,backref=true,linkcolor=black]{hyperref}
\hypersetup{
  pdfauthor = {Jean-Daniel Fekete and Romain Primet},
  pdftitle = {Progressive Analytics:
    A Computation Paradigm for Exploratory Data Analysis},
  pdfsubject = {},
  pdfkeywords = {},
  colorlinks=true,
  linkcolor= black,
  citecolor= black,
  pageanchor=true,
  urlcolor = black,
  plainpages = false,
  linktocpage
}
\usepackage{xcolor}

\renewcommand{\manuscriptnotetxt}{}

\onlineid{0}
\vgtccategory{Research}
\vgtcinsertpkg

\newcommand{\pv}{ProgressiVis\xspace}
\newcommand{\module}[1]{\texttt{#1}}
\newcommand{\add}[1]{#1}

\title{Progressive Analytics:\\
  A Computation Paradigm for Exploratory Data Analysis}

\author{Jean-Daniel Fekete and Romain Primet}
\authorfooter{
\item
 Jean-Daniel Fekete is with INRIA. E-mail: Jean-Daniel.Fekete@inria.fr
\item
 Romain Primet is with INRIA. E-mail: Romain.Primet@inria.fr
}

\shortauthortitle{Fekete \& Primet: Progressive Analytics}

\abstract{ Exploring data requires a fast feedback loop from the
  analyst to the system, with a latency below about 10 seconds because
  of human cognitive limitations.  When data becomes large or analysis
  becomes complex, sequential computations can no longer be completed
  in a few seconds and data exploration is severely hampered.  This
  article describes a novel computation paradigm called
  \emph{Progressive Computation for Data Analysis} or more concisely
  \emph{Progressive Analytics}, that brings at the programming
  language level a low-latency guarantee by performing computations in
  a progressive fashion.  Moving this progressive computation at the
  language level relieves the programmer of exploratory data analysis
  systems from implementing the whole analytics pipeline in a
  progressive way from scratch, streamlining the implementation of
  scalable exploratory data analysis systems.  This article describes the
  new paradigm through a prototype implementation called \pv, and explains
  the requirements it implies through examples.
} 


\CCScatlist{
 \CCScat{K.6.1}{Management of Computing and Information Systems}%
{Project and People Management}{Life Cycle};
 \CCScat{K.7.m}{The Computing Profession}{Miscellaneous}{Ethics}
}

\begin{document}

\maketitle

\section{Introduction}

Data analysis systems delivering their results in a progressive manner
have become more popular in the recent years
\cite{Pive,Vizdom,TMIPR12,Glueck:2014:DEP:2556288.2557195,Hellerstein:1997:OA:253260.253291,Pezzotti:2015,progressiveva,MDSteer}, but their
implementation has, so far, been ad-hoc.  This article describes a
novel computation paradigm we call
\emph{Progressive Analytics}, aimed
at improving the scalability of exploratory data analysis systems.  To
concretely explain implementation issues, we also describe a
proof-of-concept implementation of our paradigm: a Python toolkit
called \pv.  Progressive Analytics, and therefore \pv, is designed to
support the implementation of exploratory analysis systems that
compute and return results in a progressive way at the language and
library level.  Moreover, Progressive Analytics provides mechanisms to
support user interactions while the system runs, allowing filtering
data, changing parameters, and even steering algorithms: all the
underlying mechanisms needed to support visualization and visual
analytics.

Existing analytics system rely on sequential computations to deliver
their results, i.e. if a function $f$ operates on results of a
function $g$, it has to wait for $g$ to finish before starting to run.
The main issue addressed by the Progressive Analytics computation
paradigm relates 
to one intrinsic limitation of sequential systems: their latency.
With current systems, computation time can grow without bounds when
exploring large datasets or when using complex analysis methods on the
datasets.  However, when performing exploratory analysis, humans need
a latency below about 10 seconds to remain focused and use their
short-term memory efficiently \cite{Miller:1968:RTM:1476589.1476628,
  Shneiderman:1984:RTD:2514.2517}.  Sequential systems cannot
guarantee such a latency so there is always a large-enough dataset or
complex-enough computation that leads sequential systems to exceed this
limit and become unusable for exploration.  Distributing and
parallelizing computation increases the overall speed, but does not
change the sequential paradigm: the time to obtain the results of a
computation has no guaranteed bounds unless the programmer implements
these bounds in the software somehow.

In contrast, the Progressive Analytics computation paradigm is designed to always
produce partial results of computations under user-specified bounds,
allowing the system to comply with the human cognitive constraints.  A
major divergence from sequential systems is that Progressive Analytics
systems do not deliver the whole computation at once; instead, they
generate a sequence of partial results that converge to the final
results.  Meanwhile, users can interact with Progressive Analytics
systems to change parameters of the computation, instead of
waiting to completion to change them.  Analysts can also
make decisions before the end of the computation if they feel
confident about it. These early decisions are very important in exploratory
settings~\cite{Pezzotti:2015}.

Due to their appealing properties, over the last decade, many
experimental systems delivering progressive results have been
implemented, demonstrated, and described in scientific publications,
over the whole range of topics related to analytics:
databases~\cite{Hellerstein:1997:OA:253260.253291}, computations and
machine-learning~\cite{Pezzotti:2015,MDSteer},
visualization~\cite{Pive,OpeningTheBlackBox,progressiveva,MDSteer}, and user
interfaces~\cite{Vizdom,TMIPR12,Glueck:2014:DEP:2556288.2557195}.
However, all these systems rely on ad-hoc software architectures, none
of them claiming to be designed in a modular fashion or reusable.
Today, creating a new progressive system implies reimplementing almost
completely all the components from data management to user interfaces;
an overly expensive endeavor.  To address this software architecture
problem so as to streamline the development of good-quality
progressive systems, we have designed the \pv toolkit as an
experimental platform to build progressive
steerable exploratory systems in Python.

The contributions developed in this article are: 
\begin{itemize}[nosep]
\item a formal description of the Progressive Analytics computation paradigm to
  clarify the main properties needed for a system to claim
  ``progressiveness'',
\item a description of the \pv experimental implementation of
  the paradigm to show its practicality and the challenges ahead.
\end{itemize}

The article is organized as follows: the background section details
the cognitive reasons why progressive systems are needed for
exploratory analytics, defines formally what we call the ``Progressive
Computation'' paradigm, and how previous work has addressed the issues
raised by the Progressive Analytics computation paradigm.  We then provide a
detailed description of the \pv toolkit, through the implementation of
example applications and a discussion\footnote{This article
supersedes the workshop note~\cite{fekete:hal-01202901}.}.

\section{Background}

Computer users are already familiar with progressive behaviors: the
progressive loading of web pages, where text appears initially with a
rough layout, followed by images also progressively loaded, until the
web page layout tidies-up to adopt its final appearance.  Specific
mechanisms have been built in web browsers, image file formats, and
the HTML document format to support this progressive behavior and
prevent users from waiting idly while the web page and its components
get downloaded.  Progressive Analytics can be seen as an extension of
these behaviors for exploratory analytics where the guaranteed
progression of computation and the ability to interact at any time to
steer the analysis are an integral part of the analyst's activity.

\subsection{Latency, Cognitive Constraints and Exploration}
\label{sec:latency}

When exploring data, analysts operate in a tight loop where they plan
for one analysis---expecting to obtain new information about the data,
launch the appropriate algorithm, and wait for the result to confirm
or infirm the expectation.  The result arrives with some
latency.
Multiple studies have shown that humans have
perceptual and cognitive limitations related to that latency.
Miller's seminal article on the effect of computer response time
distinguishes several orders of magnitudes for latency
summarized by Nielsen in~\cite[Chapter
5]{nielsen1994usability} and \cite{limits3} (the latency names are
added by us):\\
\begin{itshape}
\noindent\textbf{Continuity Preserving Latency} \SI{0.1}{\second} is about the
  limit for having the user feel that the system is reacting
  instantaneously, meaning that no special feedback is necessary
  except to display the result.\\
\noindent\textbf{Flow Preserving Latency} \SI{1.0}{\second} is about the limit
  for the user's flow of thought to stay uninterrupted, even though
  the user will notice the delay.  Normally, no special feedback is
  necessary during delays of more than \SI{0.1}{\second} but less than
  \SI{1.0}{\second}, but the user does lose the feeling of operating
  directly on the data.\\
\noindent\textbf{Attention Preserving Latency} \SI{10}{\second} is about the
  limit for keeping the user's attention focused on the dialogue.  For
  longer delays, users will want to perform other tasks while waiting
  for the computer to finish, so they should be given feedback
  indicating when the computer expects to be done. Feedback during the
  delay is especially important if the response time is likely to be
  highly variable, since users will then not know what to expect.
\end{itshape}

Exploratory data analysis is concerned primarily with analytical tasks
where preserving attention and flow are always essential. For visual
analytics where graphical interaction is required, \emph{continuity
  preserving latency} is essential during \emph{direct
  manipulation}~\cite{DirectManipulation} phases.

Liu and Heer \cite{Liu2014} have studied the effect of
\emph{continuity preserving latency} during direct manipulation.  They
showed that ``initial exposure to delays can negatively impact overall
performance even when the delay is removed in a later session''.
Therefore, a latency under 0.1 second is essential during direct
manipulation and should be supported by an analytics system using
visualization.

Regarding the user's attention, Miller also argues that ``The
graphical response should begin within 2 seconds and certainly be
completed within 10 seconds if the user is to maintain thought
continuity in an ongoing task.''
\cite{Miller:1968:RTM:1476589.1476628}.  Shneiderman provides similar
bounds: ``These long delays may or may not increase error rates in the
range 3--15 seconds, but they will probably increase error rates above
15 seconds [...]. Dissatisfaction grows with longer response times.''
\cite{Shneiderman:1984:RTD:2514.2517}.

Waiting for task completion while keeping user's attention can be
improved using techniques called \emph{Time Engineering} or \emph{Time
  Design}~\cite{seow2008designing}.  In particular, several techniques
have been used to improve the progress bar to lower the
frustration caused by waiting, and provide a better estimate of the
waiting
time~\cite{Harrison:2010:FPB:1753326.1753556,Myers:1985:IPP:317456.317459}.
Even if some techniques only allow increasing the response time
without raising too much frustration, others such as modern progress
bars provide some information while waiting for an operation to
complete.  Indeed, the goal of \cite{seow2008designing} is to ``Build
Applications, Websites, and Software Solutions that Feel Faster, More
Efficient, and More Considerate of Users' Time!''.  In that sense,
Progressive Analytics is certainly a \emph{Time Engineering}
technique.
To achieve this goal, few strategies are possible:
\begin{enumerate}[nosep]
\item Increasing the computing speed, but that approach has well-known
  physical limits due to the speed of light both for computing and for
  transmitting data;
\item Pre-computing indexes or aggregates to speed-up the most
  common analytical queries.  However, it is almost impossible to
  pre-compute enough indexes/aggregates to guarantee an acceptable
  latency during open-ended explorations.  Furthermore, this
  pre-computation can be very long~\cite{Nanocubes}, and delays the
  moment when data can be explored;
\item Leaving the exploration to some machine intelligence that does
  not suffer from the human cognitive constraints.  However,
  intelligent systems are not yet able to replace human analysts, they
  are mainly used to provide hints and suggestions on interesting
  properties of data by computing e.g.\ correlations and searching
  remarkable properties in data;
\item Use progressive analytics, as argued in this article.
\end{enumerate}
Note that these strategies are by no means mutually exclusive; they can
and should be combined eventually.

\subsection{Definition}\label{sec:definition}

The word \emph{progressive} has been used in many articles but, to our
knowledge, never formally defined.  Let $f$ be a function with
parameters $P={p_1,\ldots,p_n}$ and yielding a result value $r$.  In
mathematics, the value $r$ calculated by $f$ does not involve any
time:
\[
f(P) \mapsto r
\]
In computer science, a computation takes a time $t$, uses an amount of
memory $m$, starts with a machine state $S$, and modifies this state
that becomes $S'$.  Also, for analytics, we distinguish the parameters
$P$ of the function from a set of input \emph{data tables}
$D={d_1,\ldots,d_m}$ that the function takes as input, and the ones it
returns as output $R = {r_1, \ldots, r_o}$. Our mathematical function
$f$ becomes a computation function $F$:
\begin{equation}\label{eq:tf}
F(S, P, D) \mapsto (S', R, t, m)
\end{equation}

For example, if $F$ is a function computing a clustering with the
k-means algorithm (e.g.\ \cite{KMeans}), it will take as
parameters (P) the number of clusters $n$, a tolerance for the
convergence of the algorithm, and possible many others depending on
the actual implementation.  It will also take one data table as input
(D), and return two data tables as output (R): the $n$ cluster
centroids, and a data table associating each input data item to its
cluster.

The \emph{progressive computation} of $F$, noted $F_p$, is a function
with three properties:
\begin{enumerate}[nosep]
\item When called repeatedly on $D_i$, a growing subset of $D$, it
  returns a sequence of partial results 
  $R_i$, each result being computed with duration time $t_i$ (\autoref{eq:prog});
\item If $q$ is the desired amount of time between two consecutive
  partial results (the \emph{quantum}), $t_i \le q$;
\item The results $R_i$ will converge to $R$ (\autoref{eq:conv}).
\end{enumerate}
We will call $F$ an \emph{eager} function, and $F_p$ a
\emph{progressive} function.

\begin{equation}\label{eq:prog}
\begin{array}{rlr}
F_p(S_1,q,P,D_{1}) & \mapsto & (S_2, R_1,t_1, m_1) \\
&\cdots & \\
F_p(S_z,q,P,D_{z}) & \mapsto & (S_{z+1}, R_z, t_z, m_z) 
\end{array}
\end{equation}

Formally, the growth of the data tables can be modeled with a
partition $P(D)$ of $D$ into $z$ non-intersecting subsets (chunks):
$P(D) = [d_1, \ldots, d_z]$. $F_p$ is therefore called with $D_k =
\bigcup_{j = 1,k} d_j$.  Note that $D_k$ does now have to grow, i.e.\
some $d_j$ can be empty.

If analysts use $F_p$ directly for their computation, $q$ should be
below \SI{10}{\second} according to the previous section. If $F_p$ is used as part
of a sequence of operations \emph{seq}, its quantum should be set so
that the time to compute \emph{seq} is below \SI{10}{\second}.  This time
constraint implies that progressive functions are \emph{real-time},
but our \SI{10}{\second} cognitive quantum, even when split to deal with sequences,
is much larger than the fractions of a second that computer-science
traditionally expect for real-time constraints.  Also, note that our
definition does not set any constraint on the memory usage $m$ of
$F_p$.  We will contrast this with other related concepts in the next
section.

We also expect that our partial results will improve all along the
computation so we define a convergence criteria for the progressive
computation of $F_p$: 
\begin{equation}\label{eq:conv}
\lim_{j \mapsto \infty} R_j = R
\end{equation}

If we define a distance $\rho(R_j,R_k)$ to measure the amount of
changes between two results, we would like our partial results to
always improve, i.e. $\rho(R, R_{j+1}) \le \rho(R,R_j)$, but, in
practice, we cannot guarantee it.  Instead, for \autoref{eq:conv} to
hold, all we need is the Cauchy convergence criteria for a sequence,
which states that, after some step $N$, the results converge.  
We would like the convergence to happen as quickly as possible
  ($N$ to be as small as possible) but this will depend on the
function $F$, the parameters $p$, the data tables $D$, the computation
method that we use, and the distance function $\rho$ that we pick.
\add{For example, using simulated annealing, the early solutions are
  often made worse on purpose before they improve.}  Also, the
convergence criteria in \autoref{eq:conv} can be too strong and there
may be some error between the exact direct computation of $F$ and its
progressive computation.  Note that the issue of errors in computation
is not specific to our progressive functions.

From the definitions above, we can see that progressive functions can
be composed just like eager ones because we know from
calculus that, if $f$ is continuous at $b$ and $\lim_{x \mapsto
  a}g(x) = b$, then:
\begin{equation}\label{eq:comp}
\lim_{x\rightarrow a} f(g(x)) = f(\lim_{x\rightarrow a}g(x)) = f(b).
\end{equation}

Practically, in analytics, there are two main causes for long
computations: 1) input/output time, and 2) computation time. If $g$ is
a progressive function that loads a large dataset $x$ 
and $f$ computes some derived value from it, then \autoref{eq:comp}
insures that $f(g(x))$ will converge to the final value progressively
as the file is being loaded ($D_k$ grows).  Conversely, if
$g$ computes a complex value derived from $x$ (e.g.\ a
multidimensional scaling over a 1 million rows dataset $x$ already
loaded so $D_k$ does not grow), then $f(g(x))$ will also
converge when the iteration converges.  Though the mathematical
foundations are sound, practical choices should be made carefully
since they will affect how useful the progressive results will be.  As
stated by Stolper et al.~\cite{progressiveva}: ``This method
[progressive visual analytics] is based on the idea that analytic
algorithms can be designed to produce \emph{semantically meaningful
  partial results} during execution.''

\subsection{Comparison with Related Concepts}

Several concepts related to progressiveness exist in analytics, such
as \emph{online}, \emph{streaming}, and \emph{iterative}.
Using the same formalism as in
\autoref{sec:definition}, we clarify their meaning and outline the
similarities and differences to our progressive functions.

\emph{Online}~\cite{online-survey} is related to the way the data
tables $D$ are made available to the function $F$.  Informally, an
online function $F_o$ will provide partial results quickly when given
data tables filled incrementally.  Online algorithms are popular in
machine learning and analytics in general, in particular to handle
dynamic data.
When $F_o$ is called iteratively with the growing subsets $D_k$, it
performs its computation efficiently:
\[\begin{array}{rlr}
F_o(S_1,P,D_1) & \mapsto & (S_2, R_1,t_1, m_1) \\
&\cdots & \\
F_o(S_z, P, D_z) & \mapsto & (S_{z+1}, R_z, t_z, m_z)
\end{array} \]
Online algorithms guarantee that the total time to compute $R$
($\sum_{j=1,z}t_j$) is much shorter than the total time to compute
each step independently (restarting the process with $S_j = S_1$).  For most
online algorithm, the total time  is assumed to be independent of the
partitioning $P$.  Just like progressive functions, online
functions imply convergence (\autoref{eq:conv}).
Contrary to progressive functions, online functions offer no guarantee
that $t_j \le q$.

Online algorithms can be made progressive if there is a guarantee that
feeding the data tables in small-enough chunks will not exceed a
desired quantum.  Then, an online function $F_o$ can be made
progressive by making sure that it will be fed with data tables in
small-enough chunks through a function $F_s$ that will split its
inputs when necessary: $F_p = F_s \circ F_o$.  The challenge is then
to find the best size of the chunks to match the time constraint.
Feeding the data tables one row at a time as many times as necessary to
fill-up the quantum is a possible strategy but usually implies a huge
overhead compared to feeding larger chunks at a time.  If the online
algorithm cannot process its input under a quantum, it cannot be
transformed into a progressive algorithm through the splitting
function $F_s$ alone.

\emph{Streaming}~\cite{streaming-survey} is related to memory usage
and time constraints.  According to Wikipedia: ``streaming algorithms
are algorithms [\ldots] in which the input is presented as a sequence
of items and can be examined in only a few passes (typically just
one). These algorithms have limited memory available to them (much
less than the input size) and also limited processing time per item.''
Given these constraints, streaming algorithms can sometimes compute
rough approximations of the theoretical results.  Streaming algorithms
are progressive by design and can be used in progressive systems,
although the quality of the results they provide can be a practical
problem.  Also, streaming algorithms are usually difficult to design
and implement; relying only on streaming algorithms to implement an
analytic infrastructure would fit the theoretical needs but severely
limit the range of possible applications that could be built.  For
example the Spark streaming machine-learning library~\cite{spark-ml}
provides 86 classes compared to the 245 provided by the scikit-learn
machine-learning library~\cite{scikit-learn}.  In comparison to
progressive functions, streaming functions have limited memory,
implying constraints on $m_i$ in \autoref{eq:prog}.

An \emph{Iterative}~(e.g.\ \cite{iterative}) method is defined in
Wikipedia as: ``a mathematical procedure that generates a sequence of
improving approximate solutions for a class of problems.'' An
iterative method performs its computation through some internal
iteration and can usually provide meaningful partial results $R_i$
after each iteration, although some extra work might be required to
obtain the partial result in an externally usable form.  Iterative
methods have a behavior similar to \autoref{eq:prog}, except they do
not take chunks but the whole dataset $D$ at each iteration.  They also
comply with the convergence of \autoref{eq:conv}.  However, being
iterative does not imply any bounds in the execution time $t_i$ of
each iteration.  If the iterations are always shorter than a specified
quantum, they can become progressive for that quantum.

\medskip

The family of progressive algorithms is the only one explicitly aiming
at user-oriented properties and guarantees.  All the other methods are
oriented towards fulfilling machine-oriented properties, not
guaranteed to be suitable for data exploration.

\subsection{Analytics and Machine-Learning}

Analytics and Machine-Learning have been concerned with online and
streaming methods for a long time, even if data exploration has rarely
been the main rationale for these topics.  Additionally, many methods
related to exploratory data visualization at scale have been pursued.

Many well-known statistical measures can be computed online or on
streams with little adaptation, including average (mean), variance,
standard deviation, and many linear models.  There are online and
streaming variants for all the families of machine-learning
algorithms, as described by M\"uhlbacher et
al.~\cite{OpeningTheBlackBox}: classification, regression, clustering,
dimensionality reduction,
and preprocessing methods (e.g.\ mean removal and variance scaling).
However, few libraries or toolkits provide a consistent software
engineering abstraction to use the progressive algorithms, with the
exception of the Spark Machine-Learning library~\cite{spark-ml} that
provides several streaming implementations to use with the Spark
cluster programming environment~\cite{Spark}.  In most cases, the APIs
to use these algorithms are ad-hoc and very diverse.

In addition to algorithms that are natively online or streaming,
substantial work has been invested into adapting incremental
algorithms to become progressive.  Stolper et al.~\cite{progressiveva}
report on how they adapted the SPAM algorithm~\cite{spam} to meet the
progressive requirements.  M\"uhlbacher et
al.~\cite{OpeningTheBlackBox} mentions 4 strategies to increase ``user
involvement'', their particular interpretation of progressiveness, and
list 12 algorithms that can be made progressive with different levels
of integration.
Directly related to progressive analytics, Pezzotti et
al. \cite{Pezzotti:2015} present a full progressive pipeline for the
t-SNE multidimensional technique.
This line of articles shows that a large family of costly iterative
projection methods can be adapted to become progressive and even
\emph{steerable}.

Algorithm steering is defined as ``interactive control over a
computational process during execution'' by Mulder et
al.~\cite{steering}. 
Progressive methods are \emph{steerable} if the user can control the
computation while the system progressively processes the rest of the
dataset.  Controlling means either changing some control parameters,
or specifying interactively that a selection of the data is more
important to the user, and that the system should deliver faster and
more accurate progressive results for that selection. Work described
in~\cite{OpeningTheBlackBox,Pezzotti:2015,progressiveva,MDSteer}
support steering.

\subsection{Visual Analytics}

Visual Analytics has been interested by two issues related to
progressive analytics: system infrastructures and HCI/cognitive
issues.

On the system infrastructure side, the closest system to ours is
PIVE~\cite{Pive}, 
and, to a lesser degree, \emph{In-Situ} visualization
systems~\cite{In-Situ}.  PIVE~\cite{Pive} allows monitoring the
progress of iterative algorithms running and sometimes even steer them
using a different approach than ours.  It instruments the iterative
algorithms to communicate with a visualization and interaction thread
running in parallel to the main execution thread.  The instrumentation
is in charge of sending the right parameters to the visualization and
interaction threads.
Using this approach, users create their
program as before; there is almost no impact on the implementation,
except that instrumentation comes afterwards and need to prepare the
data for sending to the visualization and interaction thread.  Some
instructions can also be sent in return from the visualization and
user-interface to steer algorithms, so the logic needs to be inserted
in the original algorithm.  This approach is also very close to
\emph{In-Situ} visualization~\cite{In-Situ}, a paradigm used mostly in
high-performance computing where monitoring the progression of
algorithms is very important but cannot involve too much data transfer
or program interruption because the algorithms are implemented to use
the parallel computing infrastructure in the best possible way, and
the data is just too large and dynamic to be moved.  In-situ
visualization consists of instrumenting the simulation code to allow
as much exploration as possible in the computed data without
disturbing the computation too much.  Using these monitoring
approaches, the system remains algorithm driven and not data or
analyst driven.  While this is the goal of In-situ visualization or
algorithm visualization where visual exploration is performed over the
data generated by the algorithm/simulation, general data exploration
may require more flexibility, such as trying several algorithms and
parameters to select an effective combination, or other operations
that do not consider the algorithm as being fixed upfront.

Stolper et al. \cite{progressiveva} enumerate seven important design
goals for progressive visual analytics applications.  These goals are
related to the exploration process when performed using a progressive
system. ``First, analytics components should be designed to:
\begin{enumerate}[nosep]
\item Provide increasingly meaningful partial results as the algorithm
executes
\item Allow users to focus the algorithm to subspaces of interest
\item Allow users to ignore irrelevant subspaces
\end{enumerate}
Visualizations should be designed to:
\begin{enumerate}[nosep,resume]
\item Minimize distractions by not changing views excessively
\item Provide cues to indicate where new results have been found by
  analytics
\item Support an on-demand refresh when analysts are ready to explore
  the latest results
\item Provide an interface for users to specify where analytics should
  focus, as well as the portions of the problem space that should be
  ignored''
\end{enumerate}

We have discussed Goal \#1 in the beginning of the Background section.
The other goals are related to capabilities that progressive systems
should offer, and more cognitive and HCI aspects of progressive
systems.  An infrastructure supporting the implementation of
progressive systems should provide the mechanisms to reach the listed
goals.  Unfortunately, the article does not provide enough
implementation details to compare it to our paradigm.

Finally, M\"uhlbacher et al.~\cite{OpeningTheBlackBox} describe
strategies to convert algorithms to become progressive and steerable
for a deeper \emph{user involvement}.  From a user's perspective, the
progressiveness is part of what they call the \emph{feedback}
properties, and steering is part of the \emph{control} properties.
The listed properties are essential for progressive systems and should
be implemented in the lower layers of progressive systems, such as
progress feedback, cancellation, prioritization, etc.  They also
describe 4 possible strategies to achieve their desirable properties:
1) Data subsetting, 2) Complexity selection, 3) Divide and combine,
and 4) Dependent subdivision.  All of these strategies can be
implemented in our progressive programming computation paradigm.

\subsection{Visualization} 

Progressive algorithms have been used in visualization for a long
time, in particular force-directed methods (e.g.\ \cite{tulip,d3}) for
graph layout, bubble charts, and tag clouds.  Force-directed methods
are costly and can cause latency so showing the layout being computed
is a sensible feedback.  However, to our knowledge, no study has been
performed to assess the usefulness of these animated progressive
layouts, which can be insightful or distracting.  While this strategy
is by far the most popular, it does not control the latency well; fast
systems such as Tulip~\cite{tulip} will show changes at high speed,
which can grab user's attention; slower systems may show changes every
few seconds.  Finally, some force-directed systems run continuously to
allow direct manipulation of the layout: they are therefore
progressive and steerable.

In terms of software engineering, progressive layout is usually
implemented in visualization systems using the standard drawing
pipeline: each time a new layout is available (an iteration of the
algorithm), the ``redraw'' method is called.  This implementation has
no implications for other parts of the visualization pipeline because
the drawing stage is the end of the pipeline.

Recently, Schulz et al. introduced a model for what they call
``Incremental visualization''~\cite{IncrVis}, which fits exactly our
progressive computation paradigm at the visualization level.  An incremental
visualization will receive data in chunks and update the
visualizations accordingly.  Their model also supports computational
steering.  They list the two following challenges, reiterated by
\cite{OpeningTheBlackBox}: 1) authoring incremental visualizations,
choosing what to show, how to show it, with trade-offs in complexity,
update time, and quality; 2) providing awareness of the
progression/unfinished status.

Although their model is compatible with ours, the authors mostly
mention issues related to the visualization and interaction stage, not
about the analytical components; furthermore, the latency 
  is controlled by the user.  Our model generalizes their approach
by unifying the visualization level and the rest of the pipeline,
controls the latency automatically, and provides support at the
language and library level.

\subsection{HCI}

HCI has addressed the problem of latency in very generic ways, in
particular with studies on progress
bars~\cite{Harrison:2010:FPB:1753326.1753556,
  Myers:1985:IPP:317456.317459}, and Time Design and Engineering
\cite{seow2008designing}.  All this body of work shows that some
feedback should be provided to users when the system's latency raises
above a few seconds.  The more information can be provided, the
better.  Progress bars allow users to monitor the progression of a
timely operation, but also to allocate time to other activities while
the computation progresses, with some assessment of the time
available, or giving-up on a too long task.  While the information
conveyed by progress bars is low, progressive analytics conveys useful
information over time.  Fisher et al.~\cite{TMIPR12} studied effects
of progressive visualizations with professional analysts with visual
feedback on the confidence intervals of the running results.  Their
study shows that running averages computed progressively from a
database provide useful information, allowing early decisions but some
challenges remain to interpret the confidence intervals.  On a similar
line, Glueck et al.~\cite{Glueck:2014:DEP:2556288.2557195} studied the
effect of progressive loading of time-series data. They crafted a
specific file format allowing progressive loading on time-series, and
asked subjects to find particular features at three levels of details.
They showed that progressive loading allowed these subjects to perform
these tasks faster than the non-progressive setting, especially for
coarse features that get loaded earlier.

\subsection{Databases}

Joseph Hellerstein is a precursor of progressive analytics in the
field of databases. With his colleagues, he demonstrated the
usefulness of a proof-of-concept
\cite{Hellerstein:1997:OA:253260.253291} that delivers query results
in a progressive way.  He has inspired new research in HCI and
visualization~\cite{TMIPR12}.  However, the idea has not yet found its
way to commercial solutions, maybe due to the lack of standard
software libraries and programming paradigms for connecting to
progressive databases.  Yet, the database research community is aware
of the issues raised by data exploration and are trying to address
them~\cite{RGDD11}.

The Blink DB~\cite{BlinkDB} database tries to overcome the latency
problem by providing bounded response times (or bounded error) on very
large databases.  However, it returns only one answer, not a
progressive one.  Still, Fisher et al.~\cite{TMIPR12} warns about two
issues raised by progressive exploration of databases: \emph{outlier
  values} and \emph{table joins}.  Also, most database-related work is
concerned with error bounds while returning progressive results.
Although these bounds are very useful for assessing the quality of
results, computing useful bounds might be harder or impossible for
complex computations used by data analysts.  DeLine et
al.~\cite{Tempe} have proposed a progressive database system coupled
with a streaming language to perform live computations and
visualizations.  The database is progressive, but the language is not
meant for general analytical computation.

While distributed storage and computing platforms are designed for
throughput, scalability, and resilience; not for low latency.
Recent research work has started to focus on latency, with e.g.\
Latency-Aware Scheduling \cite{Li:2015:SSA:2809974.2809979}, but
mostly for computing complete (not progressive) results.  The
dominant paradigm related to progressiveness remains streaming
\cite{spark-ml,DStreams}.  Interaction and steering are not yet
mentioned in these bodies of work.

\subsection{Summary of Features for Progressive System}
\label{sec:features}

In this section, we have gathered the following features that a
progressive analytics language and library should support:
\begin{enumerate}[label={F\arabic*:},ref={F\arabic*},nosep]
\item\label{f1} provide increasingly meaningful partial results as the algorithm
  executes~\cite{OpeningTheBlackBox,progressiveva},
\item\label{f1bis} provide feedback about the aliveness, absolute and
  relative progress of the computation~\cite{OpeningTheBlackBox},
\item\label{f1ter} provide control to cancel or prioritize
  computations~\cite{OpeningTheBlackBox},
\item\label{f2} guarantee time bounds in the delivery of progressive
  results,
\item\label{f3} allow the manipulation of progressive values, computations on
  them, and composition, similar to other first-class objects,
\item\label{f4} allow interactive steering either by modifying analysis
  parameters, or by influencing more deeply analytic component
  computations at run time~\cite{OpeningTheBlackBox},
\item\label{f5} allow general complex computations typical of exploratory
  analytics workflows, starting small and evolving by analyses
  successes, failures, and parameter-space exploration.
\end{enumerate}

These features are
meant to complement the features described for the user interface
\cite{progressiveva} or the visualization components \cite{IncrVis}.

\section{The \pv Toolkit}

\pv is an experimental implementation of the Progressive Analytics
computation paradigm in the Python language.  We chose Python
because it is among the most popular languages for Data Science,
providing high-performance libraries for a wide variety of general and
domain-specific analytics \cite{diakopoulosdata}.  \pv relies on a
well-known ``stack'' of core libraries, such as
\emph{NumPy}~\cite{numpy} for numerical computation including
high-performance linear algebra, \emph{SciPy}~\cite{scipy} for more
general scientific numerical computations, \emph{Pandas}~\cite{Pandas}
for high-performance data tables known as \emph{Data Frames} to model
data and perform statistical operations, and
scikit-learn~\cite{scikit-learn} for machine-learning.  We have tried
to stick to using the standard Python libraries, components and idioms
to assess the impact of our paradigm shift on the use of standard
libraries.  We report and comment on the main issues we encountered,
but to date, none has been blocking or serious.

\pv relies on a kernel that defines the base \emph{Module} class, a
\emph{scheduler}, and several supporting classes and mechanisms.  A
web server embedded in \pv is used to deliver visualization and manage
interaction through a web browser, but analysts can also use \pv
from within the standard Python interactive shells.

At a high-level, users create and connect modules that run
progressively in a background thread managed by a
\emph{scheduler}. \autoref{lst:heatmap} shows a simple program for
visualizing a Heatmap of the pick-up locations of the New York City
yellow taxis\footnote{The dataset is available at
  \url{http://www.nyc.gov/html/tlc/html/about/trip\_record\_data.shtml}}
in a progressive way.  Visualizing the results is done by opening a
web browser on a local port.

\lstset{language=Python}
\begin{lstlisting}[float,style=myPy,caption={Visualizing the Heatmap of
  a large data table with \pv; results appear on a web browser.},label=lst:heatmap]
from progressivis.stats import Histogram2D, Min, Max
from progressivis.io import CSVLoader
from progressivis.vis import Heatmap

csv_module = CSVLoader('yellow_tripdata_*.csv.bz2') # load many compressed CSV files
min_module = Min() # computes the min value of each column
min_module.input.df = csv_module.output.df(*@\label{line:dfmin}@*)
max_module = Max() # computes the max value of each column
max_module.input.df = csv_module.output.df(*@\label{line:dfmax}@*)
histogram2d=Histogram2D('pickup_longitude',(*@\label{line:params}@*) # compute a 2d histogram
                        'pickup_latitude',
                        xbins=512, ybins=512)
histogram2d.input.df = csv_module.output.df(*@\label{line:dfh}@*)
histogram2d.input.min = min_module.output.df(*@\label{line:mindf}@*)
histogram2d.input.max = max_module.output.df
heatmap = Heatmap() # computes the Heatmap
heatmap.input.array = histogram2d.output.df
\end{lstlisting}

In this example, the CSV module loads the New York taxi trips data
files, builds a 2D histogram to aggregate the number of pick-up
locations on a $512 \times 512$ array, and visualizes the results as a
Heatmap.  The 2D histogram needs to know the bounds (min values and
max values) of the 2 dimensions it aggregates to know the location of
its bins.  These bounds are computed by the modules
\module{min\_module} and \module{max\_module}.  While this program is
very simple, a standard eager execution needs hours to complete: just
decompressing the data files stored on a local solid-state disk for
the year 2015 (146 million lines, \SI{22}{\giga\byte}) takes about
\SI{20}{\minute} on a modern laptop; loading it completely requires
about \SI{1}{\giga\byte} of memory and hours of parsing time, far more
than the \SI{10}{\second} cognitive constraints described in
\autoref{sec:latency}.  In contrast, our progressive implementation
will start showing results in \SI{5}{\second} (since it has 5 modules
with a default quantum of \SI{1}{\second} each) and improvements at
the same pace.

The \module{csv\_module} starts loading as many lines as possible
from the files during its default \SI{1}{\second} quantum.  It
stores them in a Data Frame (hence the \emph{df} name for slots). This
data will be fed to the input of the 3 modules \module{min\_module} (line
\ref{line:dfmin}), \module{max\_module} (line \ref{line:dfmax}), and
\module{histogram2d} (line \ref{line:dfh}).  This example shows
important concepts, e.g.\ modules are connected through input and output
\emph{slots}. 
The implementation of \pv requires several important and
sometimes non-standard mechanisms that we describe now:
\begin{enumerate}[nosep]
\item a Scheduler to iteratively run progressive functions,
    as specified in \autoref{eq:prog};
\item a Module class to implement progressive functions, as
    specified in \autoref{eq:tf};
\item a mechanism to manage the time-to-run in modules to
    implement the \emph{quantum};
\item a unified representation of data;
\item a mechanism to manage the changes performed by modules so that
  the subsequent modules know what happened.
\end{enumerate}

The convergence property in \autoref{eq:conv} is left to the
implementation of each module.

\subsection{Scheduler}

The scheduler is in charge of running all the modules, according to
\autoref{eq:prog}; it starts its first iteration with a run number 1,
and increments it every time it runs a module, producing a virtual
time: the \emph{run number}.  This run number is then used to compute
what needs to be updated in modules.

The scheduler maintains a table that maps virtual time to absolute
time.  The advantages of using virtual time rather than real-time are
that: 1) managing an integer is much easier than a date, and 2) the
time of a particular run is only known when it finishes whereas the
virtual time is assigned at the start of each iteration and can be
used from inside the module when it runs.

The \pv scheduler has two modes: in normal mode, it iterates over a
round-robin queue of modules, check if the next module is ready and,
if so, runs it.  If the module is not terminated, it is then pushed
back in the queue.  We describe the second mode called
\emph{interaction mode} in \autoref{sec:interaction}.

The queue is constructed when modules are added to the scheduler,
destroyed, or the connections are changed.  A topological order is
computed over the modules so they can run in dependency order, and the
queue is set with that order.  Modules can be changed at run-time,
with some care because the scheduler runs modules in its background
thread and changes can come from the main thread (the interpreter in
Python).  \pv thus provides convenience functions to run the changes
in the background thread and avoid locking the background thread for
too long.

\subsection{The Progressive Module}

Looking back at our formal definition in \autoref{eq:prog}, a
progressive function becomes a module, the parameters $P$ are
specified during the creation of the module and can be changed later
(e.g.\ the parameters of the histogram module line
\ref{line:params}). Input data $D$ are specified using input
connections, and output data through output connections.  The quantum
can be specified as a parameter and defaults to \SI{1}{\second}.  No
further information is needed to run a progressive program, but we
detail how modules are implemented to explain the low-level
mechanisms, and the responsibilities of the system/toolkit programmer
to relieve the application programmer from explicitly managing the
progressiveness of the system.  Also, while most libraries or toolkits
consider file loading as a function that is called and returns a
value, \pv needs to create a module to manage the loading in a
progressive way.

\begin{lstlisting}[float,style=myPy,caption={\pv Module class simplified},label=lst:module]
class Module(object):
    parameters = [('quantum', float, 1.0)] #  parameter of all the modules
    def __init__(id=None, scheduler=None,**kwds): # modules belong to 1 scheduler
    def create_dependent_modules(*params, **kwds): # for complex modules
    def destroy(): # delete a module and free its associated resources
    def id(): # get the identifier of this module, unique in its scheduler
    def scheduler(): # get the scheduler
    @property def parameter(): # get the module parameter dictionary
    @property def lock(): # get the module lock 
    def has_any_input(): 
    def get_input_slot(name):
    def validate_inputs(): # check if the mandatory slots are connected
    def has_any_output():
    def get_output_slot(name):
    def validate_outputs(): # check if the slot types match
    def validate(): # check if inputs and outputs are valid
    def is_visualization(): # true if this module is a visualization
    def get_visualization(): # get the name of that visualization
    def is_input(): # true if this module manages input messages
    def from_input(msg): # receives an input message
    def describe(): # prints a readable description of the module
    def to_json(short=False): # returns a dictionary describing the module state
    def get_data(name): # get a managed data by name
    @property def state(): # get the current state
    def is_ready(): # return true if the module is ready to run
    def run(run_number): # run the module for a step when ready
    def predict_step_size(duration): # predict the number of steps for a duration
    def run_step(run_number,step_size,howlong): # performs a sub-step
\end{lstlisting}

\autoref{lst:module} summarizes the main methods of the \pv Module
class.  Input data is specified through named input connections, and
output data through named output connections (feature \ref{f3}).  For
example, the ``Min'' module has one mandatory input connection called
``df'' (see line \ref{line:dfmin} of \autoref{lst:heatmap})
and one output connection called ``df'' (see line
\ref{line:mindf}). The input of one module should connect to the
output of another module and no cycle is allowed.  The output of one
module can be connected to multiple inputs of other modules, which is
a typical asymmetry in dataflow or workflow systems. Some input
connections are mandatory and others are optional.  A module cannot
run if its mandatory inputs are not connected.  A progressive system
cannot run if there is a cycle in the connections.

In addition to its data inputs and outputs, the module parameters are
available as data and can be changed if desired.  Each module has an
optional input called \module{\_param} that can be connected to a module
to control or feed the input parameters. Also, each module has an
output called \module{\_trace} that provides a data frame with running
statistics about the module, useful for monitoring or better
predicting its performance.

Two types of modules are recognized in a special way: \emph{input
  modules} and \emph{visualization modules}. An input module is meant
to have some of its state changed by an external interactive system,
so as to propagate it through the progressive system.  It is used to
implement dynamic queries and steering (feature \ref{f4})
in a progressive system and described later in
\autoref{sec:interaction}.  Visualization modules are taking care of
preparing the visualization of analytics information; they need to be
recognized by the user interface of a \pv system, and also by the
scheduler for managing interaction.

The \pv execution of a module is dependent of an internal \emph{state}
that can take the following values: 
\emph{created}, \emph{blocked}, \emph{ready}, \emph{running},
\emph{zombie}, and \emph{terminated}.
This state is used by the scheduler to decide how to process modules.
A module is ready when either its state is \emph{ready}, or when its
state is \emph{blocked} and one of its inputs has new data available.
When a module is ready, the scheduler calls its method
\verb|run(run_number)|, which implements the logic to drive the
execution of the module instance by calling the abstract method
\verb|run_step(run_number,step_size,howlong)| with two computed
parameters: \emph{step size} and \emph{howlong}.  The last is simply
the time left to run, the first is a number of internal steps to
perform before returning and will be described in the next section.
The method \verb|run_step| is where actual computations happen.  It
returns its next computation state and the number of steps actually
run. It can be re-run if its state is ready and the quantum has not
been exceeded (feature \ref{f2}). Otherwise, the control is released
from the module to let the scheduler choose another module.

\subsection{Time to Run Management in Modules}

Although the principle of allowing a module to run for a specified
quantum of time is intuitively simple, implementing this behavior is
not that simple.  In the worst case, each module could measure the
time after each instruction and stop when getting close to the
quantum, but measuring the time continuously is costly, and some
operations are much faster when done in batches than item per item.
For example, computing the max values of data frame columns uses
optimized vectorized code that is an order of magnitude faster than
when done item per item, especially with an interpreted language like
Python.  Still, the actual time will vary from one computer to
another, and depending on the computer's load.

To address this problem of run-time prediction, \pv uses a \emph{Time
  Predictor} (feature \ref{f2}).  This manager uses the trace of the
dynamic behavior of each module collected by the scheduler, to
estimate the number of internal steps each module should perform to
match its quantum.  Internal steps can be translated by the
  module into a number of data items to process (data chunking), or
  number of iterations to perform, or any combination.

For example, when reading a CSV file, the CSV loader will initially
read a small number or rows (from 800 to 1200 in our case) and measure
the time-per-item it takes using a linear regression.  After a few
internal iterations, it will have a decent approximation of the number
of rows per second that it can read for the specific dataset and
file-system or communication channel.  With this more accurate
approximation, the module can read more (or less) rows to match its
quantum effectively.  In that case, the number of steps is
  directly mapped to the data chunking.

When a module implements an algorithm that is not linear in the number
of items processed, the Time Predictor is still effective because it
only uses a few recent runs for its estimate.  Assuming the function
mapping time-to-run to number-of-steps is not chaotic, the Time
Predictor will compute a value close to the first derivative of the
processing speed (number of steps per second).  This estimation is
therefore not perfectly accurate but good enough.  In our experience,
this approach works effectively even with functions with quadratic
complexity such as the computation of pairwise distances between rows
of a data being progressively loaded.  However, these computations
will end-up blocking when data grows because they are not able to
perform even one step during their quantum.  Then, the only option for
the user is to increase the quantum, which can lead back to the
latency issue; therefore, purely quadratic algorithms should probably
be avoided in progressive computations.  To be conservative, modules
run in at least 4 sub-steps so they can control their termination time
more accurately.  When they have a quantum $q$, they call the
\verb|run_step| method with a number of steps estimated for $q/4$,
which allows for the time predictor to better adapt to changes in the
complexity or machine load.  Whereas the standard Time Predictor
manages modules by default,
it can be overridden if a module needs finer control.

\subsection{Unified Representation of Data}

We chose to use Pandas \emph{Data Frames}~\cite{Pandas} to represent
most of the data managed by \pv.  Data Frames are similar to database
tables and are popular in statistical software such as R~\cite{R}.
Most of the values exchanged by \pv modules are data frames that we
extend with a column representing the time when rows have been
updated.  Therefore, the data frames are decorated with an additional
column called \verb|_update| that contains the run number (not a
date).  

Using one column to keep track of changes in data offers limited
precision to the nature of the changes. One value can be changed, or
the whole row created or changed: the \verb|_update| column does not
tell what happened.  However, in our experience, this level of detail
is enough for most applications.  In addition, users can model their
problems with multiple data frames if they want a finer precision.

\subsection{Change Management in Modules}

When a module has an input data frame, it usually needs to know what
happened to the data since the last run.  A \emph{change manager} provides
that information, using the \verb|_update| column and additional
information that it stores in each module input slot.  This
information boils down to what data has already been processed, and
sometimes a buffer of data to process.

From that information, each module can get the set of created rows,
the set of updated rows, and the set of deleted rows. In addition,
if some rows have been buffered for further management in the
previous runs, they are still in the buffer ready to be managed.

With these potentially three sets, each module can decide how best to
pursue its computation.  When only new data is available, module that
manages their input incrementally can just continue to do so.  For
example, a module computing the min value of each column on a data
table can just continue updating these values for the incoming rows.
If previous data has been modified or deleted, the module should
either restart its computation from scratch, or try to repair it if it
maintains a backup of previous values.  In the case of simple modules,
such as the one performing min computation, recomputing the values is
fast (typically millions of values per second).  

\begin{lstlisting}[float,style=myPy,caption={Simplified \texttt{run\_step} method of the ``Min'' Module},label=lst:min]
def run_step(self,run_number,step_size,howlong):
    dfslot = self.get_input_slot('df')(*@\label{line:begchange}@*) # access the input slot named 'df'
    dfslot.update(run_number)(*@\label{line:update}@*) # compute the changes since last run
    if dfslot.has_updated() or dfslot.has_deleted():
        ... # when anything has changed, just restart from the beginning
    indices = dfslot.next_created(step_size)(*@\label{line:indices}@*) # get the newly created rows
    steps = len(indices)
    input_df = dfslot.data() # access the data table
    op = input_df.loc[indices,self._columns].min() # apply 'min' on the new rows
    op['_update'] = run_number # set the _update column to the current time
    if self._df is None: # First time called
        self._df = pd.DataFrame([op], index=[run_number])
    else:
        op = pd.concat([last_row(self._df), op], # updating the max with the last
                       axis=1).min(axis=1)       # value and the new one
        op['_update'] = run_number               # Fix UPDATE_COLUMNS after 'min'
        self._df.loc[run_number] = op            # store at the index 'run_number'
    return {'next_state': dfslot.next_state(), 'steps_run': steps}
\end{lstlisting}

For example, \autoref{lst:min} shows how the ``Min'' module computes
progressively the minimum values over a set of specified columns.
Line \ref{line:begchange} manages the changes
that occurred in the input slot called ``df''.  The changes are
computed for the current ``run number'' by the method
\verb|update(run_number)| on line \ref{line:update}.  If the input
data frame has rows with updated values, or deleted values, then the
minimum cannot be recomputed incrementally and the computation is just
reset to be restarted from the beginning of the data frame.
Otherwise, the overall minimum values are trivially computed from the
current running values stored in the instance variable
\verb|self._df|, and the minimum values computed over the newly
created rows, returned in the ``indices'' variable line
\ref{line:indices}.

Changes can also happen in the ``columns'' dimension, when new
attributes are appearing due to e.g.\ the progression of loading.  The
most common case is when a computation needs to create modules for
each of the columns of an input data table; this happens usually the
first time the \verb|run_step| method of the module is run.  For
example, the visualization module ``Histograms'' visualizes all the
columns of a data table as 1-dimensional histograms.  It creates or
deletes the histograms when columns change.

\subsection{Progressive Computation and Repair Strategies}

Many strategies are possible to react to changes in the data.
Algorithmic modules such as k-means can actually incorporate changed
values in a simple way without keeping track of old values, since the
iterative computation of k-means will eventually ``fix'' the
clustering.  Other modules such as the histograms can only restart
from the beginning when values are changed or deleted.
Basically, each module needs to decide how to react to value changes,
and the best strategy depends on the semantic of the module. 

A particular case of value change is when a control parameter is
modified, usually interactively.  As discussed in the previous
section, these parameters can be used to tune the behavior of a
module.
However, when e.g.\ the viewport of a 2D histogram is changed, the
histogram module should restart the computation from scratch.  A more
interesting case occurs when changing the distance function for a
multidimensional projection such as MDS; the distance computation
module should restart from scratch, but the MDS module can start its
computation with the actual 2D positions already computed so, from a
user's perspective, the MDS will morph from the old configuration to
the new one, helping users to follow the changes.

\subsection{Interaction Mode}
\label{sec:interaction}

The first mode of the scheduler was using a straightforward
``round-robin'' strategy.  The second mode of the scheduler is meant
to react when an input module has been interacted with, for
  steering the module or for dynamic queries, through the method
\verb|from_input(msg)|. \autoref{sec:steering} shows how steering
  can be implemented.  To manage dynamic queries, for example to
filter-out some range in a data frame column, \pv provides a module
type called ``Variable'' that maintains a data frame meant to be
interactively changed (touched).  Its module method
\verb|from_input(msg)| receives a dictionary of values and sets its
data frame accordingly.  In our example, it can receive from our web
server a message such as:\\
{\small
  \verb|{'query': '-74.20 < pickup_longitude < -73.1'}|}\\
a query to select taxi pickup coordinates inside the New York City
area.  This query is then fed to a ``Select'' module that can be
connected to the simple Heatmap visualization pipeline in
\autoref{lst:heatmap}.  When notified by an input module that it
  has been touched, the scheduler:
\begin{enumerate}[nosep]
\item selects a subset of modules to run, 
\item accelerates the processing to react in less than
  \SI{100}{\milli\second}, so as to comply to the \emph{continuity
    preserving} latency (feature \ref{f4}),
\item runs the subset of modules only once in interaction mode,
\item reverts to normal mode unless another input has been touched.
\end{enumerate}

The subset of modules to run (the reaction subset) is computed from
the modules dependency graph. It spans from the touched input modules
to all the visualizations impacted, leaving out the modules that are
not on a path between touched inputs and impacted visualizations.
Heavy computational modules can test if the scheduler is in
interaction mode to lighten their computations.

For each input module, the scheduler computes and maintains a
\emph{reachability} set using the following algorithm:
\begin{enumerate}[nosep]
\item Compute a transitive closure of all the modules using the
  Dijkstra algorithm; we collect for each module its
  \emph{reachability set}: the list of directly and indirectly
  connected modules.
\item Collect all the modules that contain no visualization in
  their reachability set; we know that when an input is triggered,
  there is no reason to run them because they do not lead to updating
  a visualization.
\item Remove from the reachability sets of the input modules the
  reachability sets of the module collected in step 2.
\end{enumerate}
When inputs are touched, the scheduler gathers all the input modules
that have been touched, computes the union of their reachability sets
and adds them to subset of modules to run to complete the interaction.
Only these modules are considered by the scheduler when iterating over
the queue.  Once they are run, they are discarded from the temporary
set.
To ensure that the loop is run in less than \SI{100}{\milli\second},
we adopt a conservative strategy: the scheduler counts the number of
modules $n$ to run and temporarily assigns a quantum $q=0.100/n$ to
each of them.  Note that this capability of \pv is, to our knowledge,
unique and very specifically tailored for interaction, direct
manipulation, and visual analytics.

\section{Example Applications}

As explained in \autoref{sec:features}, typical exploratory
visualizations start with small pipelines that are expanded according
to results obtained in the exploration process (feature \ref{f5}).  We
show here how some of these steps can unfold using a scripting
interface; the same instructions can be packaged with a complete GUI.

When a multidimensional dataset is loaded, simple descriptive
statistics can be computed on the columns, and histograms can be
showed for all the columns using the ``Histograms'' visualization.
Exploring the New York Taxi datasets is best done using a scatterplot
visualization, as listed in \autoref{lst:scatterplot}. 

\lstset{language=Python}
\begin{lstlisting}[float,style=myPy,caption={Scatterplot visualization
    of the New York Taxi Dataset; results appear on a web browser.},label=lst:scatterplot]
from progressivis.io import CSVLoader
from progressivis.vis import Scatterplot

csv_module = CSVLoader('yellow_tripdata_*.csv.bz2')
scatterplot = ScatterPlot('pickup_longitude', 'pickup_latitude', scheduler=s)
scatterplot.create_dependent_modules(csv,'df')(*@\label{line:deps}@*)
\end{lstlisting}

\begin{figure}\centering
\includegraphics[width=0.9\columnwidth]{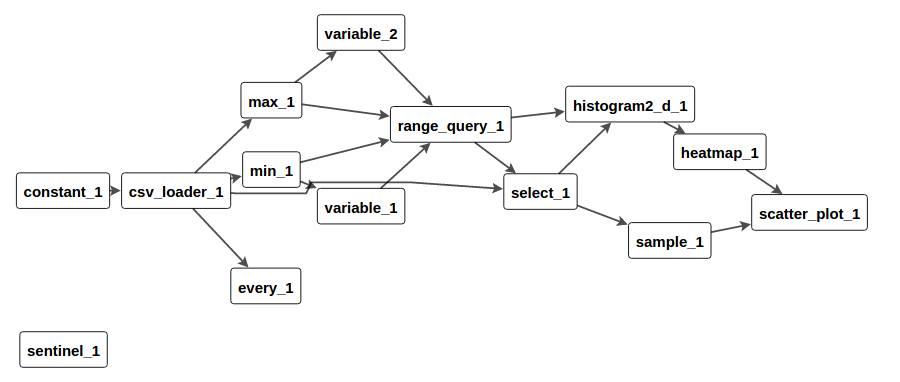}
\caption{Dependency graph of the \pv modules created in
  \autoref{lst:scatterplot}.}
\label{fig:scatterplotgraph}
\end{figure}

\autoref{fig:scatterplotgraph} shows the graph dependency view of the
pipeline, labeled with the internal identifiers of the modules.  In
addition to the 2 modules created explicitly (\module{csv\_loader\_1}
and \module{scatter\_plot\_1}), we see many others created by the
method \module{create\_dependent\_modules}. Two modules \module{min}
and \module{max} progressively compute the absolute minimum and
maximum values over all the columns of the dataset.  The values are
sent to the \module{range\_query} module that will generate a range
query to the \module{select} module, according to two
\module{variable} modules.  These modules are \emph{input} modules
that maintain a data frame of user-specified minimum and maximum
values to implement the range selection.  The two variable modules are
dependent of the \module{min} or \module{max} modules---as visualized
by a link in the dependency graph---through an input slot called
``like'' that allows the variable modules to create a data frame with
the same schema as their input slots.  Initially, these variable
values are undefined so the \module{range\_query} module generates no
query and the \module{select} module does not filter any row of the
data frame. 

The range-query module provides a set of range-sliders for our web-based
interfaces.
As explained in
\autoref{sec:interaction}, the \module{range\_query} module produces a
data frame with only one column called ``query'' that contains a
select expression such as \verb|-74.20 < pickup_longitude < -73.1|.

\begin{figure}
\includegraphics[width=\columnwidth]{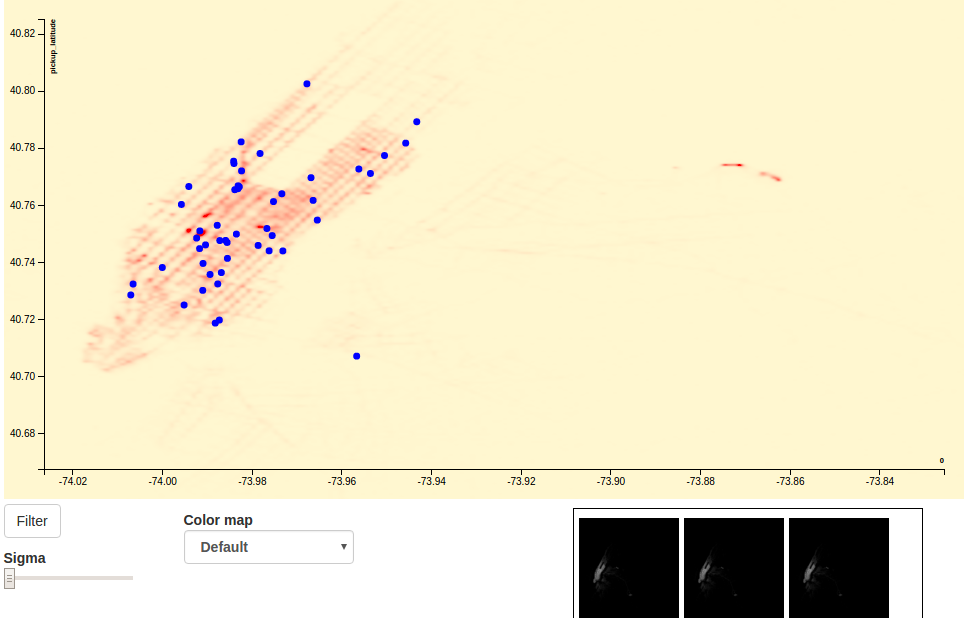}
\caption{Scatterplot over a Heatmap visualized according to
  \autoref{lst:scatterplot}, showing the pickup locations of New York
  yellow cabs (Manhattan and the airports).  New controls appear below
  the visualization: a filter button to focus on the visible view, a
  slider to change the bandwidth of the Heatmap PDF kernel on the
  bottom, a selection list to choose a colormap below, and a history
  of previous Heatmaps to explore the progression.}
\label{fig:scatterplot}
\end{figure}

The visualization is visible on \autoref{fig:scatterplot}, using a
Heatmap with points overlaid.  The points in that case are random
samples; if a function of interest is available, the points can be
chosen as the most interesting, or the most typical.  The scatterplot
can be explored using pan and zoom, but the resolution of the Heatmap
($512\times512$ bins by default) is not changed unless the ``Filter''
button is selected. This button also changes the \module{range\_query}
module to select the visible part of the display and hence increase
the resolution of the Heatmap by allocating all the bins to the area
of interest.

\subsection{Progressive K-Means}

It is important to be able to adapt well-known data analysis methods
to \pv so as to provide all the building block to implement powerful
systems (feature \ref{f5}).  K-means clustering is a classical scalable method that is
well suited to a progressive implementation.  The classical
algorithm~\cite{KMeans} is iterative.
The variant called Mini Batch k-means~\cite{MinibatchKMeans} provides
the control needed by \pv to bound the time of each run
by using a stochastic update during the iterations
instead of a deterministic one.  The two algorithms converge to very
close results~\cite{MinibatchKMeans}.  This section explains some
technical issues with adapting an iterative algorithm for \pv.  
Several iterative algorithms can be made progressive by using
stochastic iterations so the principle is generalizable, although the
actual implementations might differ greatly.

The classical k-means algorithm is not bounded in time because, for a
data table of length $n$:
\begin{enumerate}[nosep]
\item its inner loop
  runs until a numerical convergence condition is
  verified, not a bounded number of iterations;
\item the iterative computation of the label
  associated with each point (the closest cluster centroid)
  takes time proportional to $n*k$, which will
  grow when the data table grows;
\item the centroid computations will also run in time proportional to $n$.
\end{enumerate}

The Mini Batch k-means algorithm splits the whole algorithm in mini
batches, typically $100$ points, and ``fixes'' the cluster centroids
with data from the new batches.  The outer loop is run a specified
number of times or stops when the convergence criteria is reached.
Integrating that algorithm into \pv requires a few steps:
\begin{enumerate}[nosep]
\item deciding what algorithm unit will be mapped to one \emph{step};
\item removing the unbounded iterations from the algorithm so that it
  complies with the \verb|run_step()| method;
\item handling the input and outputs correctly, in particular to avoid
  generating too much work for dependent modules.
\end{enumerate}

\lstset{language=Python}
\begin{lstlisting}[float,style=myPy,caption={Core of the Mini Batch K-means Module in \pv},label=lst:nbkmeans]
class MBKMeans(DataFrameModule):
    ...
    def run_step(self, run_number, step_size, howlong):
        dfslot = self.get_input_slot('df')
        dfslot.update(run_number)
        ...
        indices = dfslot.next_created(step_size)
        steps = indices_len(indices)
        input_df = dfslot.data()
        X = self.filter_columns(input_df, indices).values
        batch_size = self.mbk.batch_size or 100
        for batch in gen_batches(steps, batch_size):
            self.mbk.partial_fit(X[batch])

        self._df = pd.DataFrame(self.mbk.cluster_centers_, columns=self.columns)
        self._df[self.UPDATE_COLUMN] = run_number
        return {'next_step': dfslot.next_state(), 'steps_run': steps}
\end{lstlisting}

We integrated the scikit-learn~\cite{scikit-learn} implementation of
the algorithm in \pv in a relatively straightforward way (see
\autoref{lst:nbkmeans}).  We then extended the implementation in two
ways: we added a steering capability, and we created a mechanism to
compute the point labels without triggering too many recomputations.

\subsubsection{Steerable K-Means}\label{sec:steering}

A \pv module needs to declare that it handles interaction by
returning ``True'' from the method \verb|is_input()| and by
defining the method \verb|from_input(msg)|.  In our case, we want
to be able to change interactively the cluster centroids, because the
k-means algorithm can become trapped in local minima and the
progressive variant is only doing one pass over the input points,
never reconsidering the initial choices it has computed for the first
input rows.  The local minima problem exists for normal k-means, but
the second problem only happens with progressive (and online) k-means.
See \autoref{lst:inter} for the simplified code to implement steering.

\lstset{language=Python}
\begin{lstlisting}[float,style=myPy,caption={Adding Interaction and Steering support in \autoref{lst:nbkmeans}},label=lst:inter]
def from_input(self, msg):
    logger.info('Received message %s', msg)
    for c in msg:
        self.set_centroid(c, msg[c])
    self.reset(init=self.mbk.cluster_centers_)

def set_centroid(self, c, values):
    c = int(c)
    centroids = self._df
    run_number = self.scheduler().for_input(self)
    centroids.loc[c, self.columns] = values
    centroids.loc[c, self.UPDATE_COLUMN] = run_number
    self.mbk.cluster_centers_[c] = centroids.loc[c, self.columns]
\end{lstlisting}

Our scatterplot module supports dragging the cluster centroids and
feeding them back to the k-means module.  When a cluster center or
several are moved and the new position sent to the module, a
progressive recomputation is triggered with the new cluster centroids
specified in the initialization.  The module is responsible for
providing cognitively understandable
feedback~\cite{progressiveva,OpeningTheBlackBox}.  We tried multiple
implementations of the steering and settled on re-running the
algorithm from the beginning, using the interactively specified
centroids as the initial configuration.  Alternatively, we could have
just continued the algorithm with the new centroids, but the quality
seemed worse.  These kinds of decisions have to be made for each
steerable module, and can sometimes be left to the user for finer
control.

\subsubsection{Stabilizing Labeling}

The k-means algorithm computes cluster positions, but can also compute
the ``labeling'' of the input points: which cluster center is the
closest to each point.  This labeling should in principle be performed
every time cluster centers change.  Iterative algorithms such as
k-means improve solutions continuously: the cluster centroids are
updated at each iteration when new points arrive.  Using the raw
progressive computation paradigm, all the modules dependent on the cluster
centroids would need to work all the time to take into account the
changes and so the point labeling would be recomputed all the time.
To avoid too many recomputations, a simple strategy consists in
delaying the propagation until some amount of change is measured.  \pv
provides a module \module{select\_delta} that receives a data table as
input, considered as a table with multidimensional data, and
propagates it as output, but only for rows that changed more than a
threshold.  The labeling module can then be connected to that
\module{select\_delta} module to recompute the labels when enough
change occurred. 

\section{Discussion}

\pv shows that typical exploratory analytics computations can be
performed in a progressive way with either a script-style or a
GUI-style interface.  Currently \pv is organized in 9 sub-packages
(\module{core}, \module{io}, \module{stats}, \module{metrics},
\module{cluster}, \module{vis}, \module{server}, and
\module{dataset}). Its core is made of 27 internal classes and
provides 36 modules with regression tests. It can be downloaded at
\url{https://github.com/jdfekete/progressivis}.

The implementation choices in \pv are the simplest we could find to
prove the feasibility of progressive analytics; we will improve these
choices later when we gain more experience about progressive systems,
the pitfalls of our choices, and better alternatives.  For example,
our scheduler does not attempt at optimizing the overall work of
modules in a pipeline: a loader can be relatively slow whereas all the
following modules perform their computations very quickly on the
arriving data, never using their whole quantum.  A smart scheduler
could allocate more time to the loader and less to the other modules
so as to comply with a total expected iteration time.  Currently, our
scheduler just obeys to the modules quantum, except in the
transient interactive mode.  Since the quantum is a tunable
parameter of each module, we rely on analysts to change it
interactively at runtime according to their preference.

\pv focuses on the implementation of the mechanisms needed to run
progressive programs.  User requirements from HCI such as providing
progress-bars or convergence/quality information (feature \ref{f1bis})
are not implemented yet or not consistently, mainly because more work
is needed to understand the whole set of extra information required to
fulfill the HCI and cognitive requirements, and the way the
information should be provided.

\pv is scalable, but remains within the capabilities and limits of its
implementation in Python.  Still, it delivers progressive results at a
human-processable speed, using effectively the main memory and CPU.
With modern computer architectures using multiple cache levels and
solid-state disks, the virtual memory of a laptop can reach
\SI{1}{\tera\byte} and the processing speed depends mostly on cache
prediction; see \cite{BMW08} for a discussion about main memory and
what it has become in the era of multiple cache levels.  \pv can
perform progressive computations efficiently, but there will always be
an overhead for progressiveness due to the requirement to generate
intermediate results.  Still we were surprised when running regression
tests of \pv against the standard Python/NumPy implementations to
realize that allocating memory in a progressive manner seems to suit
the operating system better than by large chunks that sometimes make
it ``freeze'' for a minute.  More tests are required to understand if
that behavior is exceptional or frequent.

The current Python implementation suffers from shortcomings of the
standard libraries.  For example, the SciPy library offers functions
to compute the $k$ largest eigenvector/eigenvalues of a matrix using
an iterative solver, essential for
e.g.\ computing a PCA progressively.  However, these functions are
black-boxes and do not expose their iterations.  Offering
lower-level functions to control the iterations would spare \pv from
reimplementing them.  The same is true for algorithms provided
by other Python libraries such as Pandas and scikit-learn.
We understand that libraries and new computation paradigms co-evolve so,
to address these issues in the future, we need to advertise \pv to
explain our requirements and convince library authors to better
support the needs of progressive systems.

\section{Conclusion and Future Work}

This article presented the progressive analytics computation paradigm
and its experimental implementation \pv.  We introduced the paradigm,
defined it formally and contrasted it with related concepts.
We then described the implementation of the experimental \pv toolkit,
meant to support the development of progressive systems; it can
be used to explore data using a command-line interface and
web-based interactive visualizations.  Through examples, we
showed how to explore data with progressive analytics.

In the near future, we want to extend our implementation to become a
fully usable system and toolkit for scalable analytics.  In
particular, we want to provide a rich set of progressive analytical
modules by devising strategies to adapt families of algorithms to our
progressive computation paradigm.  For example, while implementing
computation modules, we found multiple possible strategies to turn
online and iterative algorithms progressive, using hybrid solutions
such as running the main algorithm in a parallel thread and returning
predictions between long iterations.  We also want to further explore
how standard visualization techniques can be adapted for
progressiveness.

In the forthcoming years, we will explore the multiple new research
questions raised by the progressive analytics computation paradigm on
algorithms, machine-learning, cognition, HCI, visualization, visual
analytics, databases, and distributed computing.

\acknowledgments{
This work has been conducted during my Sabbatical at NYU-Poly in the
Visualization and Data Analytics (ViDA) group, partially funded by the
Center for Data Science at NYU.  Thanks to Juliana Freire for the
fruitful feedback she provided me, to Marie Douriez for her help
in implementing and experimenting Progressive MDS, and to Andreas
M\"uller for his advises on Scikit-Learn.
}

\bibliographystyle{abbrv}
\bibliography{progressivis}
\end{document}